# A Failure Self-recovery Strategy with Balanced Energy Consumption for Wireless Ad Hoc Networks

Tie Qiu, Wei Wang, Feng Xia*, Guowei Wu, and Yu Zhou
School of Software, Dalian University of Technology, Dalian 116620, China
Email: qiutie@dlut.edu.cn; f.xia@ieee.org

*Abstract*—In energy constrained wireless sensor networks, it is significant to make full use of the limited energy and maximize the network lifetime even when facing some unexpected situation. In this paper, all sensor nodes are grouped into clusters, and for each cluster, it has a mobile cluster head to manage the whole cluster. We consider an emergent situation that one of the mobile cluster heads is broken down, and hence the whole cluster is consequently out of work. An efficient approach is proposed for recovering the failure cluster by selecting multiple static sensor nodes as the cluster heads to collect packets and transmit them to the sink node. Improved simulated annealing algorithm is utilized to achieve the uniform deployment of the cluster heads. The new cluster heads are dynamically changed in order to keep balanced energy consumption. Among the new cluster heads, packets are transmitted through multi-hop forwarding path which is cost-lowest path found by Dijkstra's algorithm. A balanced energy consumption model is provided to help find the cost-lowest path and prolong the lifetime of the network. The forwarding path is updated dynamically according to the cost of the path and residual energy of the node in that path. The experimental results show that the failure cluster is recovered and the lifetime of the cluster is prolonged.

## I.   INTRODUCTION

Wireless Sensor Networks (WSNs) are widely used in environmental monitoring, disaster relief, health care and so on. WSNs are composed of large numbers of sensor nodes which are battery-powered, with sensing and limited computation as well as communication capabilities. Their major work is to sense the environment and route data packets to the Base Station (BS) via multi-hop path [1, 2]. Due to the battery constraint, the sensor nodes collect the useful information and transmit them over a long time period [3]. And the majority of energy consumption is expended on forwarding the packets [4]. How to provide a balanced energy consumption achieving maximum extension of lifetime and improving Quality of Service (QoS) of network has become a research focus in recent years.

Large numbers of sensor nodes are grouped into clusters [5], and each cluster has a cluster head. Vupputuri et al. [1] propose an efficient way using a mobile data collector(DC) as the cluster head to collect the data from static sensor nodes. Then the DCs aggregate the data and transmit it to the BS. DCs have the capability of motion which can be controlled. Because the energy of the sensor nodes is mainly expended on transmitting the data, hence in a multi-hop network, it will consume more energy if the node is closer to the sink node [6]. Consequently, DCs should change their location dynamically according to the energy of different area. DCs will move to the area which is of high energy. DCs play an important role in the WSNs, however, there is a shortcoming that if one of the DCs is out of work, whole cluster won't work anymore. So it is necessary to focus on self-reorganization for the WSNs and balanced energy consumption in unexpected situation.

In this paper, we propose a solution of self-recovery strategy with balanced energy consumption in wireless ad hoc networks with the failure cluster which is out of work. Simultaneously, the energy consumption of the node is minimized and we can keep the original performance of the whole WSNs. When meeting the unexpected situation, the WSNs can recover the cluster by itself. The rest of the paper is organized as follows. Section II describes the related work and the problem statement. In Section III, we define the energy model, and propose an improved simulated annealing algorithm to achieve the uniform deployment, besides we use the Dijkstra's algorithm to find the cost-lowest forwarding path. In Section IV, we will discuss the simulation results. Finally, Section V concludes the paper.

## II.   RELATED WORK AND PROBLEM STATEMENT

### A.   *Related Work*

Due to the strict energy constraints of the nodes, it is extremely essential to optimize the energy consumption for WSNs. Energy consumption consists of transmission cost and aggregation cost [7]. And there is also another energy cost for transmission and reception of the packets and an energy cost depending on the distance between two nodes in transmission to balance the energy consumption [8]. In addition, the node can be set to several states, such as low-energy consumption state, transmission state, receiving state and so on. In LEICP [9], in order to prolong the lifetime of the WSNs, a fitness function is defined to balance the energy consumption in every cluster according to the residual energy and positions of nodes. In [10], the authors present a comprehensive energy model for a fully function wireless sensor network, and the model is divided into 2 parts, energy consumption due to synchronization and the energy consumption due to data transmission. Besides, the energy for empty frames and missed part are also analyzed. In this paper, we also provide an energy consumption model like [10]. However, it is important to apply the energy model in finding the cost-lowest path and the formation of the cluster. Moreover, in ad-hoc WSNs, the forwarding path should vary dynamically.

---
*Corresponding author

In order to optimize energy consumption and maximize the life time of the WSNs, balanced sensor deployment as well as cost-lowest path is found by Dijkstra's algorithm in [11] and [12]. Dijkstra's algorithm will find the cost-lowest path based on the path distance, while we propose to apply the energy model to the Dijkstra's algorithm to select the optimized path. So it is reasonable to have uniform energy consumption.

It is of great significance for the cluster heads to have a better coverage of all the sensor nodes in its cluster [13]. Recently, there are many ways to detect the sensor and maximize the coverage. In [6], the authors propose that cluster heads perform parallel particle swarm optimization to maximize the coverage matrix. In LEACH, they use simulated annealing to find the most optimized location for the cluster heads. In [14], simulated annealing is also utilized to optimize localization. In the procedure of the simulated annealing algorithm, it will select the adjacent nodes to be compared with the current selected cluster heads, and the group of nodes with the minimum distance sum which is calculated from all the other nodes to the cluster heads will be the cluster heads. Consequently, it will make a contribution to the uniform distribution of the nodes. In [15], the parallel distributed self-organization clustering protocol based on clustering architecture is proposed. And the WSNs are partitioned into many small logic zones distributed uniformly according to the geography locations of the nodes. In this paper, we select cluster head not only based on the distance, but also the number of the times for which the node has been the cluster head, because the cluster head will be updated dynamically. In [16], the method is to dynamically schedule sensors' work cycles or sleep cycles in the heterogeneous WSNs. The author used a multiple criteria decision making method to optimize the sleep scheduling process. These studies have successfully obtained good results in energy optimization and extension of the maximum lifetime of WSNs, but failure self-recovery strategy with balanced energy consumption for WSNs with multiple clusters is rarely mentioned.

*B. Problem Statement*

However, almost all recent papers assume that the nodes are all in well condition. But we cannot deny that majority of the sensor nodes are exposed in the nature, perhaps it will break down for physical reasons (e.g. components of the circuit is damaged by high temperature or water) or technical reasons (e.g. a software bug in the system), especially if the cluster head is out of work, the performance of the whole WSNs will go down. So it is important and necessary to detect and solve this problem quickly.

A WSN is divided into multiple clusters, every cluster has one Mobile Cluster Head (MCH), which will collect the packets from sensor nodes in their own cluster, and then send the packets to BS. If a MCH is out of work, packets of the whole cluster are missed. As showed in Figure 1, the red symbol is the broken down MCH. We suggest that multiple sensor nodes should be selected in that cluster, which will act as the role of the failure MCH. The new cluster heads will take responsibility of collecting the packets from member sensor nodes. Besides, among the new cluster heads, a cost-lowest forwarding path will be found to transmit the packets to the adjacent MCH or BS. So the balance of the energy consumption in the failure cluster needs to be focused on to maximize the lifetime of the whole cluster in this unexpected situation.

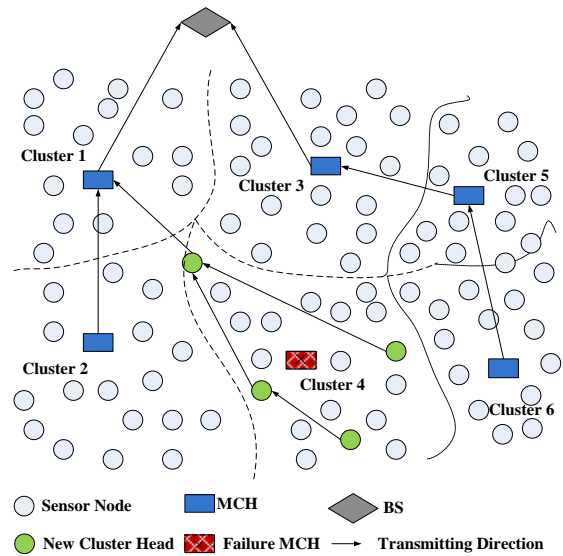

Figure 1. A WSN with failure MCH

We simply describe a WSN with failure MCH and present the approach as showed in Figure 1. The WSN is divided into six clusters (cluster 1-6). Data packets of sensor nodes are collected by Cluster head for each cluster in the WSN. If the cluster head close to the base station, the data packets will be collected directly to the base station. If the cluster head far away from the base station, data packets are sent to the base station through other cluster heads. As showed in Figure 1, MCH of cluster 4 breaks down suddenly, and the whole cluster is consequently out of work. If a sensor node is selected to substitute the breakdown MCH, the node energy will be consumed rapidly, which leads to the extremely early death of the node. This paper is aimed at re-organizing the failure cluster by balanced energy consumption of the new cluster head. The new cluster heads and the forwarding path found by Dijkstra's algorithm are constantly updated, ultimately achieving balanced and minimum energy consumption to prolong the network lifetime.

### III. MODELING AND ANALYSIS

We propose a strategy to re-organize the sensors by themselves. Our solution is to filter out the eligible sensor nodes to be cluster heads at first. The second step is to find the cost-lowest path according to Dijkstra's algorithm which is based on the energy consumption model. Finally, the path weight will be changed dynamically based on the energy model. Consequently, the forwarding path is also changing during the packets transmission. Then the whole networks can work normally. Furthermore, our contribution is to recover breakdown cluster without extra mobile cluster head. More importantly, the life time is almost the same as the original networks.

#### A. Eligibility of Sensor Nodes

First of all, the eligible sensor nodes need to be filtered out. In LEACH, it selects the eligible sensor nodes simply according to the energy, while the selection process is based on the location and distance of the nodes. But this procedure may cause that the node with high energy as well as good location (specifically, it may be relatively close to the centre of the area) may be selected as the cluster head frequently, which will have a negative effect on the energy balance in the whole cluster. As a result, we select the cluster heads according to their residual energy and the times for which they have been the cluster head. For this restriction, the sensor is eligible, only if the energy of the sensor surpasses the average energy of the whole cluster, and the number of times it has been the cluster head is less than the average times. Thus not all the sensor nodes which surpass the average energy are equal. The nodes are more likely to be selected which have never been the cluster head. It is defined as:

If $E_i > \overline{E}$ and $P_i > \overline{P}$ then the $i$-th node is eligible.

$$\overline{E} = \frac{\sum_{i=0}^{nn} E_i}{nn} \quad (1)$$

$$\overline{P} = \frac{\sum_{i=0}^{nn} P_i}{nn} \quad (2)$$

where $nn$ is the number of the sensor nodes in this cluster, $E_i$ is the residual energy of $i$-th sensor node's, $\overline{E}$ is the average energy of the whole sensors in the cluster, $P_i$ is the number of the times for which $i$-th node has been the cluster head, $\overline{P}$ is average times of the whole sensors. The nodes' energy and the eligible of the sensor nodes vary dynamically. In every round when changing the cluster heads, eligibility needs to be calculated again.

#### B. Uniform Deployment for New Cluster Heads

The uniform distribution of the cluster heads is also important to the energy balance of the nodes. We discuss how to select the cluster heads from eligible nodes to make the cluster heads' location well-distributed based on the distance and energy. Thus coverage and the energy efficiency can be provided for WSNs.

We have improved the model of simulated annealing, and we add our evaluating standard into the annealing model to optimize the process. The theory of simulating annealing originated in the physical theory, and the concept is based on the manner in which liquids freeze or metals re-crystallize in the process of annealing [14], which can be used to locate a good approximation to the global optimum of a given function in a large search area. Here we can use this theory to select a location of the cluster heads to perform the best coverage and the energy efficiency. The process is:

When the algorithm of simulated annealing starts, the sensor nodes are initialized to be a high energy level. Then multiple sensor nodes are randomly selected from the eligible sensor nodes as the initial cluster heads. The number of the cluster heads depends on the scale of the cluster, packets length, and so on [16]. Next, the optimization process will begin. At first, the cost of the initial cluster heads will be calculated. Then each cluster head's neighbor will be randomly selected as a new group of cluster heads. Neighbors' existence must be confirmed. And the cost of the new cluster head is calculated, if the cost of the new cluster heads is less than the current cluster heads, the group of new cluster heads will substitute current cluster heads in this step, otherwise current cluster head is the same. At last, previous process is repeated until the iteration is at the end. Thus the best choice of the new cluster heads is found out.

The cost is defined according to not only the distance of sensor nodes but the energy state of each one.

$$\text{cost} = \sum_{j=1}^{nn} \alpha \times MIN(D_{ij}) \times (1 + P_i / P_s + \beta / E_i)(i = 1, 2, 3 \ldots N_c) \quad (3)$$

where $\alpha$, $\beta$ are const numbers which are suitable to different specific WSNs. $D_{ij}$ is the distance between $i$-th sensor node and $j$-th sensor, the function $MIN$ will find the minimum $D_{ij}$ from other cluster heads to the $j$-th sensor node. $i$ is the sequence number of cluster heads which is from 1 to $N_c$. $N_c$ is the number of the cluster heads. $nn$ is the number of the sensor nodes, $P_j$ is the number of the times for which the $j$-th cluster head has been the cluster head, $P_s$ is the total number of the times, $E_i$ is the residual energy of *i-th* cluster head. With this energy cost model used in the simulated annealing, we can obtain a better location of the cluster heads.

**Algorithm 1:** Find Optimized Cluster Heads

---
**Begin**
  **Step 1.** Initialize $D_{ij}$, $P_i$, $P_s$, $E_i$, which is defined in equation (3);
  **Step 2.** Randomly select $N_c$ cluster heads from eligible sensor nodes. The state S is initialized, $S = \{1,2,...N_c\}$;
  **Step 3.** Calculate cost of current cluster heads *CC* by equation (3);
  **Step 4.** Set minimum cost, $MIN\_C = CC$;
  **Step 5.** Set iteration condition *ITER*(It is set to be 1000 in this paper), and set the iteration value $K=0$;
  **Step 6.** Find the cluster heads with lowest cost:
    **Step 6.1** Randomly select neighbor sensor nodes of *S* as new cluster heads;
    **Step 6.2** Calculate the cost of new cluster heads *CN*;
    **Step 6.3** Decide whether to change current cluster head to be new cluster heads
      **Step 6.3.1** According to LEACH, set $ck = 1000 \times \exp(-K/20)$;
      **Step 6.3.2** Calculate probability $P_K$ of changing the current cluster heads:
        If ($CN<MIN\_C$) $P_K = 1$
          else if ($CN==MIN\_C$) $P_K = 0$
            else $P_K = \exp(-(CN - MIN\_C)/ck))$
      **Step 6.3.3** let *rand* = a random real number in (0, 1);
      **Step 6.3.4** if ($rand < P_K$)
          Change *S* and current cluster heads with new cluster heads
          $MIN\_C = CN$
    **Step 6.4** $K=K+1$, if ($K<ITER$) go to **Step 6.1**, else iteration is finished and return current cluster heads;
**End**

---

In Algorithm 1, *CN* is the energy cost of new cluster heads found from the neighbor. *CC* is the energy cost using current cluster heads. *MIN_C* is recently the minimum cost. *ITER* is the loop times in which we predict that the best location of the cluster heads will be found. *S* is the state, namely, the neighbor nodes will be picked around *S*, and it is initialized to be current cluster heads. The algorithm presents that, If *CN* is less than *MIN_C*, new cluster heads becomes new optimum and the state *S* will be changed to new cluster heads. Otherwise, new cluster heads may still become new optimum with a non-zero probability set below. Because, if *CN* is less than *MIN_C*, $P_K$ will be set to 1, so the random number distributed between 0 and 1 must be less than $P_K$, then the current cluster heads must be replaced by new cluster heads. However, if *CN* surpasses *MIN_C*, the probability of changing state *S* is not zero, but an extremely small number.

This algorithm has a relatively high complexity. However, it is the BS's responsibility to run this algorithm to reduce the extra consumption of sensor nodes. The location information of all sensor nodes needed in the algorithm is stored in the BS in advance. It will not vary due to the static sensor nodes. The energy level of each node and the times for which sensor node has been the cluster head will be attached in the data packet to be sent to the BS. Then after the process of the BS, BS needs to return the selection result to the selected cluster heads.

Now the new cluster heads are set up. Besides, by this algorithm, each sensor nodes have acknowledged that which cluster head it will be managed by, according to the minimum distance between itself and the cluster head. Then the sensor node will begin to transmit packet to its cluster head. Each packet from the cluster head will be attached with the energy value of themselves. And all of this will be send to the BS. BS is responsible to maintain the energy of the whole breakdown cluster. Then, once the BS finds any one of the cluster heads' energy is below the threshold, it will re-select the cluster heads for the cluster using Algorithm 1. We define the threshold *Th* as:

$$Th = \alpha \cdot \overline{E}^2 \quad (4)$$

In (4), $\alpha$ is a constant number, $\overline{E}$ is the average energy of the whole cluster. The value of $\alpha$ can be changed to select the occasion of changing the cluster heads to keep a balance between the uniform energy distribution of all sensor nodes and the times of changing the cluster head, which will finally lead to a best performance.

*C. Energy Consumption Model*

There must be immense and wasteful energy consumption if all the cluster heads in the failure cluster send their packets directly to the BS or adjacent MCH which is perhaps far away. It is effective to set up a multi-hop topology for the cluster heads to save energy. We suggest each cluster head select a cost-lowest way from itself to BS or adjacent MCH through multiple other cluster heads. Dijkstra's algorithm is utilized to select the cost-lowest way, and the way weight is defined as the syntheses of distance and energy.

We divide the energy consumption into two parts for each node. One is energy consumption of transmission, the other is reception consumption. They are defined separately. The reception energy consumption is a const value *C*, and the transmission part is obtained by equation (5):

$$E_t = \alpha \cdot E + \beta \cdot b \quad (5)$$

$$E = \lambda L^2 = \lambda \cdot ((ch\_x_i - ch\_x_j)^2 + (ch\_y_i \cdot ch\_y_j)^2) \cdot b \quad (6)$$

$$E_r = C \quad (7)$$

where $E_t$ is the transmission consumption, $E$ is the cost of radio based on the distance between two cluster heads, $b$ is data size of the packet. If $b$ is a constant number, $E_t$ will be just related to the radio energy. $ch\_x$ and $ch\_y$ are the coordinate of the cluster head. $E_r$ is reception energy.

*D. Cost-lowest Path by Dijkstra's Algorithm*

As what is presented previously, among the cluster heads, it is expensive if every cluster head transmit the packets directly to the BS or the adjacent MCH. It is necessary to find the cost-lowest way to the sink node. The forwarding path is probably a one-hop directly, also perhaps a multi-hop path, which all depends on the cost of the path.

In [7], Dijkstra's algorithm has been utilized to find the cost-lowest path. However, they just take the distance between two nodes into consideration. Considering that one of the cluster heads' location is so well that all the other cluster head will select it as a vertex in the path, consequently, this cluster head will expend more energy than the other cluster heads. It will cause an unbalanced energy deployment in the cluster heads, which finally results in that this group of cluster heads need to be changed so early according to equation (4). In order to address this problem, we propose a dynamical edge weight, which varies in real time. During the transmission, the path will be changed if needed according to the real-time edge weight.

Then the edge weight value between two cluster heads is defined as:

$$EW_{i,j} = E_t \cdot (1 + \theta / E_i) \quad (i \neq j) \tag{8}$$

where $E_t$ is the transmission consumption defined in equation (5). $\theta$ is a const number. $E_i$ is the *i-th* cluster head's residual energy. $i, j$ is the number of the cluster head from *1* to $N_c$ (defined in equation (3)). Base on this edge weight model, the weight depends on $E_t$ and residual energy of the cluster head. Once the path is found, the $E_t$ is a relatively stable value, and will only be changed a little if the packet size is changed, while residual energy is changing with the procedure. As the energy goes down, the edge weight will rise up correspondingly. If one of the cluster head' energy goes extremely down, the network can sense this situation, and adjust its edge weight to a high level. Then the Dijkstra's algorithm will be utilized again to find a new path for them. Obviously, the cost-lowest way is dynamically changing to achieve the best energy saving and energy balance.

Given the network Graph *<V,E>* with edge weight, Dijkstra's algorithm can find the forwarding path with minimum cost from every node to the sink. At our present situation, V is a set of the cluster heads. E is a set of edges from any one cluster head to other cluster heads. For every edge, we have an edge weight which is calculated in equation (8) stored in a matrix.

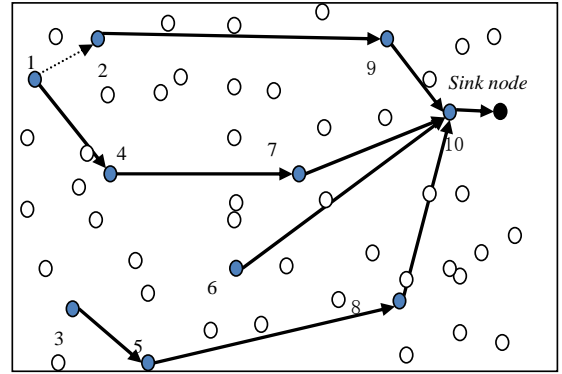

Figure 2. Cost-lowest forwarding path by Dijkstra's algorithm

At beginning, every forwarding path from the cluster head to the sink node (BS, or adjacent mobile sink node) is initialized to infinite. *S* is a set of sensor nodes whose final cost-lowest-path weights from the source cluster head have already been determined. *S* is empty initially. In the algorithm, it will repeatedly select the vertex *u* from V-S with the lowest-cost path estimated, then add it to *S*, and update the entire determined path leaving *u*. Through $N_c$ times loop, the forwarding path from every cluster head to the sink will be set up. The running time of this Dijkstra's algorithm depends on how the min-priority queue is implemented, and we achieve the running time of $O(V^2/lgV)$. This is practical to implement the min-priority queue with a binary min heap. Furthermore, for a definite time, the edge weight matrix will be updated with the energy going down, and the Dijkstra's algorithm will be called again to re-build the path.

TABLE I. INFORMATION OF NODES

| Node No. | $ch\_x$ | $ch\_y$ | $E_i$ |
|---|---|---|---|
| 1 | 9.6532 | 90.9823 | 0.9263 |
| 2 | 17.7690 | 102.5647 | 0.9256 |
| 3 | 26,1295 | 26.6744 | 0.9230 |
| 4 | 31.2210 | 68.2318 | 0.9584 |
| 5 | 40.6683 | 5.9087 | 0.9906 |
| 6 | 59.9032 | 35.6784 | 0.9756 |
| 7 | 75.3462 | 67.9984 | 0.9920 |
| 8 | 90.7640 | 8.0864 | 0.9716 |
| 9 | 99.9736 | 103.9833 | 0.9667 |
| 10 | 117.9065 | 82.9472 | 0.9575 |

Figure 2 shows one scenario of the forwarding path selecting. There are ten cluster heads which are numbered from 1 to 10, and one sink node. In Table 1, it shows the coordinates and energy level of each node. In order to find the cost-lowest path, according to the equation (8) and Algorithm 2, there are 4 paths selected which are cost-lowest and leads to a uniform deployment of energy. Our strength is not only focusing on the energy consumption but also energy deployment, which will make a contribution to prolonging the whole networks lifetime.

**Algorithm 2:** Find Cost-lowest Path

---
**Begin**
**Step 1** Initialize *SN* and *V*
      *SN* ← source nodes, such as BS or adjacent MCH
    *V* ← cluster heads
    Initialize $E_t$ and $E_i$
    Initialize *Np* which is used to record last node added in *SN*. Initialize *Np* = source node number.
**Step 2** Update edge weight value *EW* according to equation(8)
**Step 3** Iteration to find cost-lowest path
    While ($V \neq \phi$ )
      **Step 3.1** Find next node
        *u* ← find a node number from *V* which has a minimum edge weight to *Np* from *EW*
        Delete *u* from *V*.
        Add *u* to set *SN*，*SN* ← $SN \cup u$ .
        Let *p = u*
      **Step 3.2** Update former cost
        For each cluster head which is adjacent *u*, relax cost of the cluster-head
          Update the cost-lowest path.
    End While
**Step 4** Return the cost-lowest path
**End**

---

In Algorithm 2, step 1 is initialization, to set *V* and *SN*, and update the edge weight value *EW*. Step 2 is iteration to extend the forwarding path to find the cost-lowest path. In step 3, every loop we select a node from *V* which has a minimum edge weight to *p* from *EW* as node *u*, then remove it from *V*, add it into *SN*. It is calculated that with the addition of *u* whether some existent paths need to be changed, namely whether the cost is lower through node *u*. If the cost is lower through *u*, the existent paths need to be updated. This procedure is repeated until all cluster heads are added in *SN*, and *V* is empty. Thus the cost-lowest path from source to the sink node is found.

## IV. NUMERICAL CALCULATION AND EXPERIMENTAL RESULTS

In this section, we will conduct some simulation experiments to analyze the energy efficiency, energy balance and lifetime time of the re-organized cluster as well as performance. In this simulation, we focus on breakdown cluster which is managed by the failure MCH. According to our strategy, static cluster head will be selected to take responsibility to transmit the packets. During the simulation, energy of each node will be recorded to prove the uniform deployment of the energy, while total mounts of packets are calculated to test lifetime of networks.

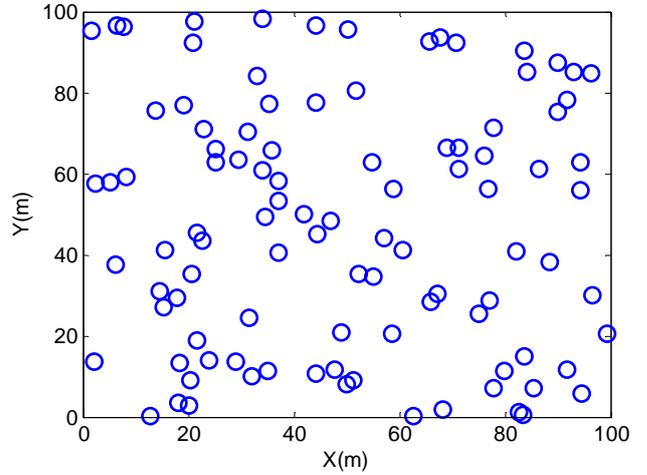

Figure 3. Sensor nodes distibution for experiment

### A. Simulation Environment

Generally, we only take the breakdown cluster into consideration. It is a topology of 100 sensor nodes (i.e. *nn*=100) and one BS which is the sink unit. We assume that all these sensor nodes are distributed in an area of $100m \times 100m$ , as showed in Figure 3. Distribution of the sensor nodes are generated by NS2. We suppose the initial energy of the sensor node is 1. We set the parameter $\alpha = 1 \times 10^{-7}$, $\beta = 1$, $N_c = 10$ in (3), namely, we will select 10 cluster heads as the cluster heads. In (4) we set $\alpha$=0.9. In (5), $\alpha = 1 \times 10^{-7}$, $\beta = 1 \times 10^{-8}$, and we set b to a const value 1K, which means that all the packets are 1K. In (6), we set $\lambda = 1$. In (7), the const reception consumption $C = 1 \times 10^{-5}$. $\theta = 1$ in (8). When the simulation ends, the total number of packets transmitted is 15263, and the total round of changing the cluster heads is 1030.

### B. Balanced Eenergy Consumption

We capture 4 moments uniformly by time in the simulation to show the residual energy of all the sensor nodes, we can observe that, at any moments, the energy of the node is at the proximity level.

Figure 4 shows the energy relationship of 100 nodes at different time points. We can see that the energy difference of any two nodes' energy is less than 20% of the whole energy. If one node's energy is at the top of all nodes' energy, it wont be the top in next period, which shows that the node of high energy is more likely to be the cluster head. And the energy consumption of each sensor node is generally similar from the beginning to the end, and the energy line is going down uniformly. It can prove that in our algorithm, the energy consumption of the nodes in the failure cluster can keep balanced.

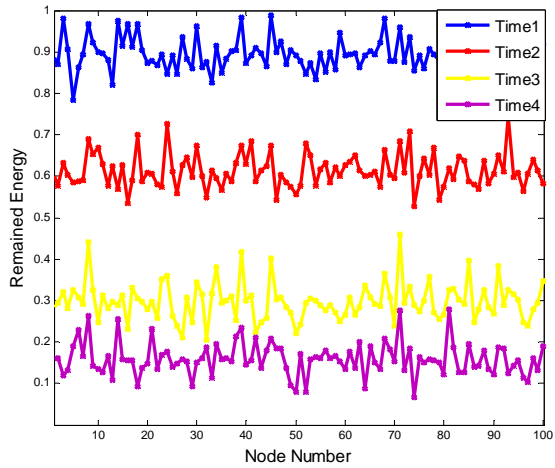

Figure 4. Remained energy of sensor nodes

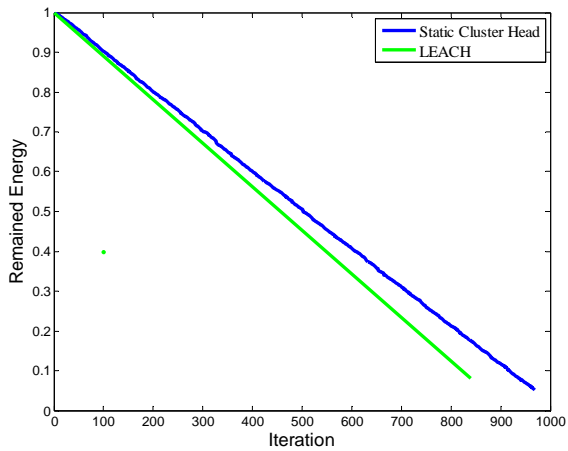

Figure 5. Average remained energy of 100 nodes

In Figure 5, the average energy of the 100 nodes is calculated in every round. It also shows that in our strategy, the average energy goes down stably, and the energy of the whole WSNs is uniformly distributed in the 100 nodes. While the lower line indicates the energy level in LEACH, the consumption of sensor nodes is larger than that in our method. The energy consumption is not uniformed due to its random selection of the member nodes. Meanwhile, the energy level of each node will fluctuate in LEACH. As for our method, using extended simulated algorithm and Dijkstra's algorithm, the consumption of the sensor nodes is uniform and minimized. It obviously reflects that lifetime of the WSNs is prolonged longer than LEACH.

Figure 6 depicts the relationship of the living nodes of the cluster. We can see that using multiple static cluster heads, 100% of the nodes are alive, which covers 90% of lifetime. However, with the MCH, at 200-th round, the nodes begin to die. But the WSNs can last a long time with MCH, because the death rate of sensor nodes is going down. Compared with the method only using DCs in [1], our method saves more energy in the whole process. Until up to the 1000th round, the nodes begin to die. In contrast, the method using only DCs starts at about the 200th round. However, in terms of the whole lifetime in our method, it is a little shorter but very close to the method using DCs.

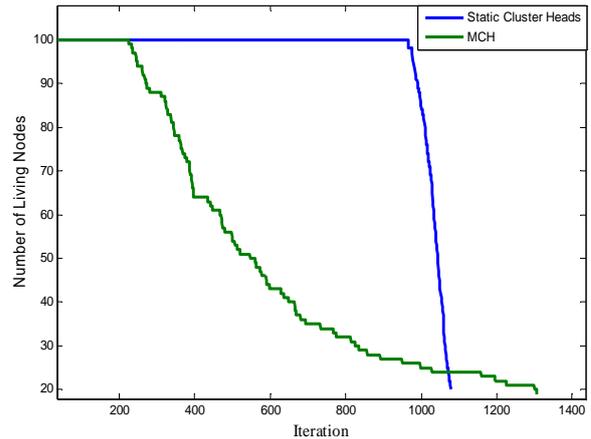

Figure 6. Number of living nodes

Figure 7 depicts the energy comparation between two situations. One situation is that cluster is recovered by multiple static cluster heads when MCH is broken down. The other is that the cluster is recovered by multiple static cluster heads at first, and then after 400 rounds, alternate MCH (AMCH) is dispatched to recover the cluster. The energy consumptions in two situations are close to each other. Due to the AMCH, the rate of the energy declines. The lifetime of WSNs is thus prolonged.

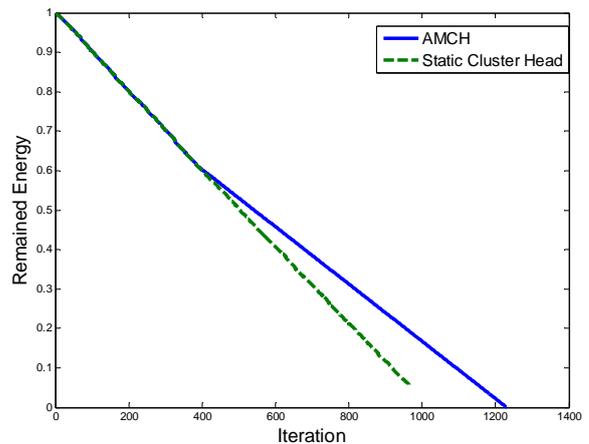

Figure 7. Remained energy

From the simulation, the contribution of our strategy is to make all the sensor nodes involved in the recovery of the cluster. All nodes have the possibility to be the cluster heads. This kind of uniform energy consumption will lead to longer lifetime. Compared with the MCH or AMCH, the performance of static cluster head is almost the same, especially in the comparison of living nodes (as shown in Figure 6) with original MCH. By using static cluster head, 100% living nodes will last longer.

## V. CONCLUSIONS

The heterogeneous WSNs consist of static sensor node, mobile cluster heads and the base station. We discuss a scenario that one of the mobile cluster heads breaks down. The sensor nodes in the failure cluster will recover themselves by selecting multiple temporary cluster heads to act as role of mobile cluster head. Improved simulated annealing algorithm is utilized to achieve the uniform deployment. An energy model is applied to keep the balanced energy consumption. We use extended Dijkstra's algorithm to find the cost-lowest path based on our energy model. Our proposed solutions achieve that the failure cluster is recovered by the multiple new static cluster heads, which can work normally. Due to the balanced energy consumption, the lifetime of the WSNs is prolonged as showed in the simulation. The strength of the solution in this paper is that the breakdown networks are recovered in a short time using the static sensor nodes, while we can also keep the well performance of the network. So the whole WSNs can live longer in complex and hostile environment. Each node can participate in the management of WSNs, so the WSNs will be stronger.

This paper discusses situation that only a small number of MCHs can not work in WSNs. If there are multiple adjacent clusters do not work, our method can also self-recover, but the efficiency will be affected. Our future work will be focused on the situation that multiple adjacent MCHs do not work normally. We will re-divide clusters to increase efficiency of the node transmission in WSNs.

ACKNOWLEDGMENT

This work is partially supported by Natural Science Foundation of China under Grant No. 60903153, the Fundamental Research Funds for the Central Universities (DUT10ZD110), and the SRF for ROCS, SEM.